\documentclass[12pt]{article}\usepackage{latexsym}
\usepackage{amsmath,,calrsfs}
\usepackage{amsfonts}
\usepackage{amssymb}
\usepackage{amscd}
\usepackage{bbm}
\usepackage{fancybox}
\usepackage{cite}
\usepackage{amsmath,amsfonts,amsbsy}
\usepackage{pstricks,pst-node}
\usepackage[small,bf,hang]{caption2}
\usepackage{graphicx}
\usepackage{epsfig}
\usepackage{psfrag}
\usepackage{comment}

\usepackage{float}

\psset{unit=1.3cm,linewidth=.5pt,radius=.2}  

\usepackage{multirow}                     
\usepackage{float}                          
\usepackage{lscape}                         
\usepackage{bm}

\newcommand{\beq}{\begin{equation}}
\newcommand{\eeq}{\end{equation}}
\newcommand{\bi}{\begin{itemize}}
\newcommand{\ei}{\end{itemize}}
\newcommand{\bt}{\begin{tabular}}
\newcommand{\et}{\end{tabular}}
\newcommand{\bc}{\begin{center}}
\newcommand{\ec}{\end{center}}

\newcommand{\be}{\begin{equation}}
\newcommand{\ee}{\end{equation}}
\newcommand{\bea}{\begin{eqnarray}}
\newcommand{\eea}{\end{eqnarray}}
\newcommand{\ba}{\begin{array}}
\newcommand{\ea}{\end{array}}

\def\bbox{{\,\lower0.9pt\vbox{\hrule \hbox{\vrule height 0.2 cm
\hskip 0.2 cm \vrule height 0.2 cm}\hrule}\,}}
\newcommand{\dsl}{\pa \kern-0.5em /}

\begin{document}

\begin{titlepage}
\begin{center}

\hfill UG-12-09\\ \hfill DAMTP-2012-10\\

\vskip 1.5cm

{\Large \bf  On  ``New Massive''  4D Gravity}

\vskip 1cm

{\bf Eric A.~Bergshoeff$^1$, J.J.~Fern\'{a}ndez-Melgarejo$^{1,2}$, Jan Rosseel$^1$ \\ 
and Paul K.~Townsend$^3$}

\vskip 25pt

{\em $^1$ \hskip -.1truecm Centre for Theoretical Physics,
University of Groningen, \\ Nijenborgh 4, 9747 AG Groningen, The
Netherlands \vskip 5pt }

{email: {\tt E.A.Bergshoeff@rug.nl, j.rosseel@rug.nl}} \\

\vskip 15pt

{\em $^2$ \hskip -.1truecm Grupo de F\'{\i}sica Te\'orica y Cosmolog\'{\i}a,\\ 
	Dept. de F\'{\i}sica, University of Murcia,\\ 
	Campus de Espinardo, E-30100 Murcia,  Spain \vskip 5pt }

{email: {\tt jj.fernandezmelgarejo@um.es}} \\

\vskip 15pt

{\em $^3$ \hskip -.1truecm Department of Applied Mathematics and Theoretical Physics,\\ Centre for Mathematical Sciences, University of Cambridge,\\
Wilberforce Road, Cambridge, CB3 0WA, U.K. \vskip 5pt }

{email: {\tt P.K.Townsend@damtp.cam.ac.uk}} \\

\end{center}

\vskip 0.5cm

\begin{center} {\bf ABSTRACT}\\[3ex]\end{center}

We construct  a  four-dimensional  (4D) gauge theory that propagates, unitarily,  the five polarization modes of a massive spin-2 particle.  These modes are described by a ``dual'' graviton gauge potential  and the Lagrangian is 4th-order in derivatives. As the construction mimics that of  3D ``new massive gravity'',  we call this 4D model   (linearized) ``new massive dual gravity''.  We analyse its massless limit, and discuss similarities to the Eddington-Schr\"odinger model.

\end{titlepage}

\newpage
%

\section{Introduction}

Einstein's theory of gravity can be viewed as an interacting theory
of massless spin-2 particles,  gravitons, but it is not excluded
that the graviton could have a small mass \cite{Vainshtein:1972sx}. This possibility has
implications for cosmology that have motivated many recent models of
``massive gravity''; see \cite{arXiv:1105.3735} for a recent review. At the
linearized level, the introduction of a mass for the spin-2 field is
straightforward, and it leads to the well-known Fierz-Pauli (FP) theory.
The difficulty is to find a consistent interacting version of this
theory\footnote{We are informed that a consistent, and ghost-free, massive gravity extension of the
four-dimensional FP-theory has now been found \cite{deRham:2010ik,deRham:2010kj}.}.

This difficulty  is much less severe in three spacetime dimensions
(3D), as is shown by  ``new massive gravity'' (NMG) \cite{Bergshoeff:2009hq} in which a
mass scale is associated with curvature-squared terms in the action.
Despite being  ``higher-derivative'', NMG has a linearized  limit  that is  equivalent to
the ghost-free 3D  FP theory,  and hence propagates  a parity-doublet of
massive spin-2 states. In addition, the Boulware-Deser ghosts \cite{Boulware:1973my} that afflict
generic interacting  higher-derivative theories  are absent \cite{deRham:2011ca}.

In this paper we explain how some of the ideas underlying the success of NMG as
a 3D massive gravity  theory can be extended to higher dimensions, in particular
four spacetime dimensions (4D). At first sight, this seems impossible because
the addition of higher-derivative interactions to the 4D Einstein-Hilbert action never
yields a ghost-free theory propagating massive gravitons \cite{Stelle:1976gc}.
However, this problem can be circumvented, at least at the linearized level, by using
a ``dual'' field representation to describe the graviton, i.e.~not the usual symmetric
tensor field but some other  gauge field  that describes the same (massive)
degrees of freedom.

The idea that the graviton might be described by a field in a ``dual'' representation of the
Lorentz group is  realised by the Eddington-Schr\"odinger (ES) model  \cite{Edd,Sch}, which has an
affine connection as the fundamental field.  We review  this model in our  concluding section
because there are some similarities to the model that we propose here, and it  shows that 
 ``exotic''  Lorentz representations can be compatible with interactions. Other ``exotic'' Lorentz representations
were explored  in \cite{Deser:1980fy}, and the topic has been  reconsidered  more recently in the context of string/M-theory dualities, 
see, e.g. \cite{West:2001as,Hull:2001iu}.  However,  what we need here is a dual representation for {\it massive} spin 2.
In $D$ spacetime dimensions, the possible spin-2 field representations are those induced by representations of
$SO(D-1)$ corresponding to two-column Young tableaux; for $D\le4$ the correspondence is one-to-one\footnote{
Any discussion of  this issue for $D>4$ depends on  what is meant by ``spin''.}. The standard
(symmetric tensor) representation has one box in each of the two columns while the dual representation has $(D-2)$ boxes
in the first column and one in the second column. The 3D case  is  special in that the ``dual'' field is again a
symmetric tensor field\footnote{Four dimensions is similarly special for massless spin-2 duality.}.

A Fierz-Pauli-type model for the 4D dual-field representation of a massive spin-2 particle was constructed by
Curtright under the rubric of  ``generalized FP theory''   \cite{Curtright:1980yk} (see also \cite{Curtright:1980yj}).
It was pointed out  relatively recently \cite{Gonzalez:2008ar} that Curtright's model can be found by an application
of ``connection-metric  duality''. Although the dual spin-2 field is not a  
gauge field in this context, there is a procedure (introduced  in \cite{Andringa:2009yc} and  explained in detail in \cite{Bergshoeff:2009tb})
for converting 3D FP-type models  into  equivalent higher-derivative gauge theories. The basic idea is to solve the
differential ``subsidiary condition''  on the FP field. A special feature of 3D is that the gauge field thus introduced
is in the {\it same} Lorentz representation as the original FP field, i.e.~a rank-$s$ symmetric tensor for spin $s$. This allows
the integration of the field equations to a  (higher-derivative) gauge invariant action, and this action  is ghost-free if it
preserves parity, which it does if  $s$ is even\footnote{This point has been clarified in \cite{Bergshoeff:2011pm}.}.
Linearized NMG may be found this way by starting with the 3D FP theory for spin 2.

A naive extension of this procedure to $D>3$ fails because the gauge field found by solving the differential constraint
on the FP field is no longer in the same Lorentz representation as the original FP field. However, if one takes the {\it dual} FP
theory as the starting point then the gauge field {\it is} in the same (dual) Lorentz representation, so one can integrate
the gauge-field equations to a (higher-derivative)  action, which  is ghost-free for some choice of overall sign because all propagated
modes are related by rotational invariance. Here we use this observation to construct a fourth-order but ghost-free
4D field theory for massive spin 2.  It is a 4D analog of the linearized 3D NMG  but with a ``dual''  field describing the
graviton, so  we shall call it  (linearized)  ``new massive dual gravity'' (NMDG). This  linear massive spin-2 model is  potentially the linearized limit of an interacting NMDG theory describing 4D massive gravity but we postpone further discussion of this issue to the conclusions. 

One issue that can be addressed in the context of a linear theory, at least partially,  is the nature of the massless limit. As is well-known, this limit is singular for FP theory because whereas the FP action describes the five polarization states of a massive spin-2 particle its massless limit coincides with the  linearized Einstein-Hilbert action, which propagates only the two modes of a massless spin-2 particle.  In the case of NMDG it is less clear what might be meant by the ``massless limit''.  One option is to consider the 4th-order term by itself.
The analogous massless limit of 3D NMG was analysed by Deser \cite{Deser:2009hb}, who found that it propagates a single mode, in contrast to NMG itself, which propagates two modes. Here we analyse  the 4D ``pure'' 4th-order model defined by the 4th-order term of NMDG. We find that it propagates three modes, one scalar and two spin-2 modes, in contrast to NMDG itself, which propagates an additional  two helicity-1 modes. 

We also show  that the massless limit of NMDG can be taken in another  way, and one that does   {\it not} lead to any discontinuity in the number of propagated modes.  In the massless model that results from this limit,  the spin-1 modes are present as Goldstone bosons which become  the helicity-1 modes of a massive graviton in NMDG itself.  This is a novel example of the St\"uckelberg  mechanism, which is essentially an affine version of the Higgs mechanism.

\section{Connection-Metric duality}

We  begin by reviewing  how connection-metric duality may be used  to construct a  ``generalised FP''  field theory for a ``dual''  spin-2 field  on a Minkowski space-time of arbitrary dimension $D$ \cite{Gonzalez:2008ar}. Then we focus on the $D=3,4$ cases. 

A convenient  starting point for this construction is the  first-order form of the Einstein-Hilbert Lagrangian
in which the vielbein $e_\mu{}^a$ and spin-connection $\omega_{\mu ab}= -\omega_{\mu ba}$
are independent fields. We then linearize about a Minkowski vacuum by writing the vielbein as
\begin{equation}
e_\mu {}^a = \delta_\mu {}^a + h_\mu{}^a\, ,
\end{equation}
and expand the action to second order in fields. After adding a 
Fierz-Pauli mass term for the perturbation $h$, we arrive at the Lagrangian
\begin{equation} \label{Laghomega}
{\cal L} = -2 h(\partial\cdot \omega) + 2 h^{\mu\nu}
\left(\partial^\alpha \omega_{\nu\mu\alpha} + \partial_\nu
\omega_\mu \right) - \omega\cdot\omega -
\omega_{\nu\alpha\rho}\,\omega^{\alpha\rho\nu} - m^2
\left(h^{\mu\nu}h_{\nu \mu} -h^2\right)\, ,
\end{equation}
where $m$ is the mass,  and
\begin{equation}
\omega_\mu = \omega^\alpha{}_{\alpha\mu}\, , \qquad h= h_\mu{}^\mu\, .
\end{equation}
It should be noted that $h_{\mu\nu}$ is a general second-order
tensor. For $m^2=0$ the action is invariant under the gauge
transformations
\begin{equation}
\delta h_{\mu\nu} = \partial_\mu \xi_\nu + \Lambda_{\mu\nu} \, ,
\qquad \delta \omega_{\mu\nu\rho} = \partial_\mu \Lambda_{\nu\rho}\, ,
\end{equation}
where $\xi_\mu$ and $\Lambda_{\mu\nu} = -\Lambda_{\nu\mu}$ are the parameters of the linearized general coordinate and
Lorentz transformations, respectively.

Eliminating the spin-connection from the action gives rise to the following Lagrangian:
\begin{equation}\label{lagFP}
{\cal L}_{h} = h^{(\delta \nu)} \varepsilon_\delta{}^{\gamma \mu}
\varepsilon_\nu{}^{\rho \alpha} \partial_\rho \partial_\gamma h_{\alpha \mu}
-m^2 (h^{\mu \nu}h_{\nu \mu} - h^2) \,.
\end{equation}
The antisymmetric part of $h_{\mu\nu}$ appears only in the mass term and is therefore an auxiliary field that can be trivially eliminated; this yields  the 
usual  spin-2 FP action for a  symmetric tensor field.  Alternatively, we can eliminate from (\ref{Laghomega}) the entire tensor $h_{\mu\nu}$ in terms of the spin-connection. After multiplication by $m^2$ this yields the ($D$-dimensional) dual Lagrangian
\begin{equation} \label{duallag}
{\cal L}^{(D)}_{dual} =  \left[K_{\mu\nu}K^{\nu\mu} -
\frac{D}{4(D-1)} K^2\right] - m^2\left(\omega\cdot\omega -
\omega_{\mu \nu \rho}\omega^{\rho \mu \nu}\right)\, ,
\end{equation}
where
\begin{equation}
K_{\mu\nu} = \partial^\alpha \omega_{\mu\nu\alpha} +
\partial_{\mu}\omega_{\nu}\, , \qquad K = K_\mu{}^\mu = 2\, \partial\cdot \omega\, .
\end{equation}
It is convenient to rewrite this in terms of a rank-$(D-1)$ tensor $U$,  defined by writing the spin-connection in the form
\begin{equation} \label{epsredef}
\omega_{\mu \alpha \beta} = \varepsilon_{\alpha \beta\nu_1\cdots \nu_{D-2}}
U^{\nu_1\cdots \nu_{D-2}}{}_\mu\, .
\end{equation}
The $U$-tensor is antisymmetric on its first $(D-2)$ indices. We now discuss separately the special cases $D=3$ and $D=4$.

\subsection{$D=3$}

In this case \eqref{epsredef} becomes
\begin{equation}
\omega_{\mu \nu \rho} = \varepsilon_{\nu \rho \alpha}U^\alpha{}_\mu \, ,
\end{equation}
where $U$ is a general second-rank  tensor. Substitution into  the kinetic term of the dual Lagrangian (\ref{duallag}) yields
\begin{equation} \label{kinterms3d}
K^{\mu \nu} K_{\nu\mu} - \frac{3}{8} K^2 = U^{\delta \nu} \varepsilon_\delta{}^{\gamma \mu} \varepsilon_\nu{}^{\rho \alpha} \partial_\rho \partial_\gamma U_{\alpha \mu} - \frac{1}{2} (\varepsilon^{\mu \nu \rho} \partial_\mu U_{\nu \rho})^2 \,.
\end{equation}
Using a Schouten identity, one can show that
\begin{equation} \label{schouten}
\frac{1}{2} (\varepsilon^{\mu \nu \rho} \partial_\mu U_{\nu \rho})^2= U^{[\delta \nu]} \varepsilon_{\delta \gamma \mu} \varepsilon_\nu{}^{\rho \alpha} \partial_\rho \partial^\gamma U_\alpha{}^\mu  \,,
\end{equation}
so that
\begin{equation}
\mathcal{L}_{\mathrm{dual}} =U^{(\delta \nu)} \varepsilon_\delta{}^{\gamma \mu} \varepsilon_\nu{}^{\rho \alpha} \partial_\rho \partial_\gamma U_{\alpha \mu} - m^2(U^{\mu \nu}U_{\nu \mu} - U^2) \,,
\end{equation}
where, here,  $U$ is the trace of the matrix $U$.
This has precisely the form of the FP Lagrangian \eqref{lagFP}. We thus find that the dual Lagrangian is  equivalent to the usual FP Lagrangian for a massive spin-2 field.

\subsection{$D=4$}

In this case \eqref{epsredef} becomes
\begin{equation}
\omega_{\mu \nu \rho} = \varepsilon_{\nu \rho}{}^{\alpha \beta} U_{\alpha \beta, \mu} \,.
\end{equation}
For clarity, we have inserted a comma to help recall which is  the antisymmetric index pair: $U_{(\alpha\beta),\mu} = 0$.  Substitution into the kinetic terms of the Lagrangian \eqref{duallag} for  $D=4$ yields
\begin{equation} \label{kinterms4d}
K^{\mu \nu} K_{\nu \mu} -\frac{1}{3} K^2= U^{\gamma \delta, \nu} \varepsilon_\nu{}^{\rho \alpha \beta} \varepsilon_{\gamma \delta}{}^{\sigma \mu} \partial_\rho \partial_\sigma U_{\alpha \beta, \mu} - \frac{1}{3}\left( \varepsilon^{\alpha \beta \gamma \delta} \partial_\alpha U_{\beta \gamma, \delta} \right)^2 \,.
\end{equation}
The 4D identity analogous to the 3D identity (\ref{schouten}) is
\begin{equation} \label{an2}
\frac{1}{3}\left( \varepsilon^{\alpha \beta \gamma \delta} \partial_\alpha U_{\beta \gamma, \delta} \right)^2=
U^{[\gamma \delta, \nu]} \varepsilon_{\gamma \delta \sigma \mu} \varepsilon_\nu{}^{\rho \alpha \beta} \partial_\rho \partial^\sigma U_{\alpha \beta,}{}^{\mu} \, .
\end{equation}
As a consequence, the totally antisymmetric part of the $U$-tensor cancels from the kinetic term and can be trivially eliminated
to yield a Lagrangian in terms of the field
\begin{equation}
T_{\mu\nu,\rho} = U_{\mu\nu,\rho} - U_{[\mu\nu,\rho]}\, ,
\end{equation}
which is a mixed-symmetry tensor with zero  totally antisymmetric part. After multiplication by a factor of $1/4$, we arrive at the dual Lagrangian
\begin{equation} \label{ldual4d}
\mathcal{L}_{\mathrm{dual}} = \frac{1}{4}T^{\gamma \delta, \nu} \varepsilon_\nu{}^{\rho \alpha \beta} \varepsilon_{\gamma \delta}{}^{\sigma \mu} \partial_\rho \partial_\sigma T_{\alpha \beta, \mu}  - \frac{m^2}{2} \left( T^{\mu \nu, \rho} T_{\mu \nu, \rho} - 2 T^\mu T_\mu \right) \,,
\end{equation}
where $T_\mu= T_{\mu\nu,\rho}\eta^{\nu\rho}$. In terms of the ``generalized Einstein tensor''
\begin{equation}\label{geneinstein}
\mathcal{G}_{\mu \nu,\rho}(T) = \frac12 \varepsilon_{\mu \nu}{}^{\alpha \beta} \varepsilon_\rho{}^{\gamma \delta \epsilon} \partial_\alpha \partial_\gamma T_{\delta \epsilon, \beta} \,,
\end{equation}
the dual Lagrangian takes the form
\begin{eqnarray} \label{FPlag}
\mathcal{L}_{\text{dual}} = \frac12  T^{\mu \nu,\rho}\, \mathcal{G}_{\mu \nu,\rho}(T)  - \frac{1}{2} m^2 \left(T^{\mu \nu,\rho} T_{\mu \nu,\rho} - 2 T^{\mu} T_{\mu} \right) \,.
\end{eqnarray}

The generalized Einstein tensor is a mixed-symmetry tensor of the same algebraic type as  $T$. 
It satisfies the  Bianchi-type identity
\begin{equation}\label{Bianchi-type}
\partial^\mu  \mathcal{G}_{\mu\nu, \rho} \equiv 0\, , \qquad \partial^\rho \mathcal{G}_{\mu\nu, \rho} \equiv 0\, . 
\end{equation}
The first of these implies the other,  as a consequence of the  fact that $\mathcal{G}_{[\mu\nu, \rho]}\equiv0$. It also 
implies that
\begin{equation}\label{Gident}
\partial^\mu \mathcal{G}_\mu = 0\, , \qquad \mathcal{G}_\mu= \mathcal{G}_{\mu\nu,\rho}\eta^{\nu\rho}\, . 
\end{equation}
Another feature of the generalized Einstein tensor is that it  defines a self-adjoint tensor differential operator acting on tensors of the same algebraic type as $T$; the field equations for $T$ are therefore
\begin{equation}
\mathcal{G}_{\mu \nu,\rho}(T) = m^2 \left(T_{\mu\nu,\rho} -2 T_{[\mu} \eta_{\nu]\rho}\right)\, .
\end{equation}
An equivalent set of equations is
\begin{eqnarray} \label{FPeqT}
& & (\Box - m^2) T_{\mu\nu,\rho} = 0 \,, \qquad T_\mu = 0\,, \qquad
 \partial^\rho\, T_{\rho\mu,\nu} = 0  \,.
\end{eqnarray}
The first of these is dynamical while the other two are ``subsidiary'' conditions, which ensure that only the five polarization modes of a spin-2 particle are propagated; this can be shown by a straightforward analysis. Also, because  $T_{[\mu\nu,\rho]}\equiv 0$, by definition,  the differential subsidiary condition implies that
\begin{equation}\label{altsub}
\partial^\rho T_{\mu\nu,\rho} =0\, .
\end{equation}

\section{New Massive Dual Gravity}

We may find a new set of higher-order equations equivalent to the second-order equations (\ref{FPeqT}) by solving the differential subsidiary condition. To do so we use the Poincar\'e lemma for differential forms in $D$-dimensional  Minkowski spacetime: for any $p$-form $P$, 
\begin{equation}\label{lemma}
\partial_\mu P^{\mu\nu_1\dots \mu_{p-1}} =0 \quad \Rightarrow \quad P^{\mu_1\dots\mu_p} = \varepsilon^{\mu_1\dots \mu_p \nu_1\dots \nu_q\rho} \partial_\rho Q_{\nu_1\dots\nu_q} \qquad (q= D-1-p)
\end{equation}
for some  $q$-form $Q$, which is defined modulo a closed $q$-form. We can apply this to any tensor that is divergence-free on some set of $p$ anti-symmetrized indices to get a dual gauge potential in which the set of $p$ indices are replaced by a set of $q$ indices. 

In the case of the  differential subsidiary condition on the $T$-tensor in (\ref{FPeqT}) we have to take into account that it also satisfies (\ref{altsub}). We therefore have to use the $D=4$ case of (\ref{lemma}) twice (once with $p=1$ and once with $p=2$). This leads to an expression for $T$ as a second-order differential operator acting on a tensor $S$  of the {\it same} algebraic type as $T$. In fact, one finds that 
\begin{equation}\label{solution}
T_{\mu\nu,\rho} = \mathcal{G}_{\mu \nu, \rho} (S) \,, 
\end{equation}
where the $S$-tensor is  defined  by this equation modulo a gauge transformation of the type
\begin{equation}\label{gaugetran}
S_{\mu\nu,\rho} \to S_{\mu\nu,\rho} + \partial_\rho \Lambda_{\mu\nu} - \partial_{[\mu}\Lambda_{\nu]\rho} +
\partial_{[\mu}\Xi_{\nu]\rho}\, ,
\end{equation}
where $\Lambda$ is an antisymmetric tensor parameter and $\Xi$ a symmetric tensor parameter. What was the differential subsidiary condition on $T$ has now become the Bianchi-type identity for $\mathcal{G}(S)$. The construction gives us the general solution of the differential subsidiary condition, so $\mathcal{G}(S)=0$ should imply that $S$ is pure gauge. This will  be verified explicitly in subsection \ref{infinite}. 

Using (\ref{solution}) in the remaining  equations of  (\ref{FPeqT})  we find the following two gauge-invariant equations for $S$:
\begin{equation} \label{NMDGeom}
(\Box - m^2) \mathcal{G}_{\mu\nu,\rho} (S)= 0 \, , \qquad  \mathcal{G}_\mu(S) = 0 \,.
\end{equation}
By construction, these equations are equivalent to the ``generalised FP'' equations (\ref{FPeqT}).
They can be derived from the following ``new massive dual gravity'' (NMDG) Lagrangian (which is gauge-invariant up to a total derivative):
\begin{equation} \label{NMDGlag}
\mathcal{L}_{\mathrm{NMDG}} =- \frac12 S^{\mu\nu,\rho}\, \mathcal{G}_{\mu\nu,\rho}(S) +
\frac{1}{2m^2} S^{\mu\nu,\rho}\, \mathcal{C}_{\mu\nu,\rho}(S)\,,
\end{equation}
where
\begin{equation} \label{Ctens}
\mathcal{C}_{\mu\nu,\rho} = \Box\, \mathcal{G}_{\mu\nu,\rho}-  \partial_\rho \partial_{[\mu}\, \mathcal{G}_{\nu]}
+ \eta_{\rho[\mu} \Box\, \mathcal{G}_{\nu]} \,.
\end{equation}
The $\mathcal{C}$-tensor is of the same algebraic type as the generalized Einstein tensor, and hence of the new mixed-symmetry tensor field $S$,  {\it except that it is traceless}. It also satisfies the same  Bianchi identities  as the generalized Einstein tensor. 

The equivalence of the 4th-order Lagrangian (\ref{NMDGlag}) to Curtright's 2nd-order ``generalized FP'' Lagrangian can be  verified directly, and this will be done in subsection \ref{aux}. Here we just explain how  the field equation of (\ref{NMDGlag}) is equivalent to the equations of (\ref{NMDGeom}). To this end, we observe that the $\mathcal{C}$-tensor defines a self-adjoint tensor operator acting on tensors of the same algebraic type as $S$. From this we deduce that the NMDG field equation is
\begin{equation}\label{NMDGequation}
m^2 \mathcal{G}_{\mu\nu,\rho}(S) - \mathcal{C}_{\mu\nu,\rho}(S) = 0 \,.
\end{equation}
Taking the trace, we deduce that $\mathcal{G}_\mu(S) = 0$, and hence from \eqref{Ctens} that
$\mathcal{C}_{\mu\nu,\rho}= \Box\, \mathcal{G}_{\mu\nu,\rho}$. Using this in (\ref{NMDGequation}), we deduce that
$\mathcal{G}_{\mu\nu,\rho}$ is annihilated by the Klein-Gordon operator, and hence that equation (\ref{NMDGequation}) is, as claimed,  equivalent to the two  equations of (\ref{NMDGeom}). 

We have now shown  that the field equations of the NMDG Lagrangian (\ref{NMDGlag}) are equivalent to the  (4D) spin-2 FP equations, despite the fact that this Lagrangian is 4th-order in derivatives.  Because all five propagating modes are in an irreducible representation of the rotation group, they will all be propagated unitarily for an appropriate choice of sign, which is the one we have chosen.  For the remainder of this section, we  explore a few features of this 4D  model that are suggested by its 3D cousin,  NMG, and we consider its infinite mass limit.

\subsection{Dimensional reduction}\label{subsec}

The analogy of  4D NMDG theory with 3D NMG may be clarified by showing how the latter follows from a truncated dimensional reduction of the former.  We will split the 4D indices  as
\begin{equation} \label{FPhd}
\mu  = \{m, z\} \,,
\end{equation}
where $z$ denotes the compactified direction. All fields are assumed to be independent of $z$. We then define
\begin{equation}
h_{mn} = S_{(m|z|,n)} \, ,
\end{equation}
and set all other components of $S$ to zero. Using this reduction/truncation in the action \eqref{NMDGlag}, we arrive at the 3D Lagrangian\begin{eqnarray}
\mathcal{L}_h & = & 2 h^{\mu\nu}\, \mathcal{G}_{\mu\nu}(h) + \frac{4}{m^2} \left[\mathcal{G}^{\mu \nu}(h) \mathcal{G}_{\mu \nu}(h) -  \frac{1}{2} \mathcal{G}(h) \mathcal{G}(h)\right] \,,
\end{eqnarray}
where the scalar $\mathcal{G}$ is the 3D Minkowski trace of the symmetric tensor $\mathcal{G}_{\mu \nu}$, which is the linearized Einstein tensor for  the symmetric tensor $h$. This is one form of  the 3D NMG Lagrangian for a massive spin-2 field.

\subsection{Auxiliary field formalism}\label{aux}

The $\mathcal{C}$ tensor appearing in the NMDG Lagrangian \eqref{NMDGlag} can be written in the form
\begin{equation}\label{altCtens}
\mathcal{C}_{\mu\nu,\rho} = 2 \,\mathcal{G}_{\mu\nu,\rho}\left(\mathcal{S}(S)\right)\, ,
\end{equation}
where the $\mathcal{S}$-tensor is an analog of the Schouten tensor:
\begin{equation}
\mathcal{S}_{\mu\nu,\rho} = \mathcal{G}_{\mu\nu,\rho} + \eta_{\rho[\mu} \mathcal{G}_{\nu]}\, .
\end{equation}
Using the self-adjointness of the operator defined by the tensor $\mathcal{G}(S)$, we may rewrite the
Lagrangian \eqref{NMDGlag} as
\begin{equation} \label{NMDGlagALT}
\mathcal{L}_{\mathrm{NMDG}} =- \frac12 S^{\mu\nu,\rho}\, \mathcal{G}_{\mu\nu,\rho}(S) +
\frac{1}{m^2} \mathcal{G}^{\mu\nu,\rho}(S)\, \mathcal{S}_{\mu\nu,\rho}(S)\,.
\end{equation}

Now consider the alternative two-derivative Lagrangian, involving an auxiliary field $f$ of the same algebraic type as the field $S$
(i.e.~$f_{ab,c} = -f_{ba,c}$ and $f_{[ab,c]}=0$):
\begin{equation} \label{NMDGaux}
{\mathcal{L}} = - \frac12 S^{\mu\nu,\rho}\, \mathcal{G}_{\mu\nu,\rho}(S) + \frac{1}{m^2} \left[ f^{\mu\nu,\rho}\, \mathcal{G}_{\mu\nu,\rho}(S) -
\frac{1}{2} \left( f^{\mu\nu,\rho}\, f_{\mu\nu,\rho} - 2 f^\mu \, f_\mu \right) \right]\,.
\end{equation}
The field equation for $f$ is
\begin{equation} \label{eomS}
f_{\mu\nu,\rho} = \mathcal{S}_{\mu\nu,\rho}(S) \, .
\end{equation}
Using this equation to eliminate the $f$-field from (\ref{NMDGaux}) we recover the NMDG action in the form (\ref{NMDGlagALT}).

The alternative NMDG Lagrangian (\ref{NMDGaux}) allows an alternative proof of  the equivalence of the linearized 4D NMG to the dual FP theory. It is simplest to first  diagonalize by setting
\begin{equation}
S_{\mu\nu,\rho} = \tilde{S}_{\mu\nu,\rho} + \frac{1}{m^2} f_{\mu\nu,\rho} \,.
\end{equation}
We then find that the Lagrangian (\ref{NMDGaux})  becomes
\begin{equation}
\mathcal{L} = -\frac12 {\tilde S}^{\mu\nu,\rho}\, \mathcal{G}_{\mu\nu,\rho}(\tilde S) + \frac{1}{2m^4} \left[ f^{\mu\nu,\rho}\, \mathcal{G}_{\mu\nu,\rho}(f) -  m^2\left(f^{\mu\nu,\rho}\, f_{\mu\nu,\rho} - 2 f^\mu\, f_\mu \right)\right] \,.
\end{equation}
The first term is just the infinite mass limit of the Lagrangian (\ref{NMDGaux}) expressed in terms of $\tilde S$. As it propagates no modes (a statement that we verify in the subsection to follow)  it  may be ignored.
The remaining terms are just those of the dual FP Lagrangian \eqref{FPlag} expressed in terms of the $f$-tensor field.
This argument is analogous to the one originally used to show  the equivalence of 3D NMG to the 3D FP theory \cite{Bergshoeff:2009hq}.

\subsection{Infinite mass limit}\label{infinite}

In the long distance limit, which is equivalent to a  limit in which the graviton mass becomes infinite, the Lagrangian of NMDG reduces to the second-order one
\begin{equation}\label{L2}
\mathcal{L}_2 = - \frac12 S^{\mu\nu,\rho}\, \mathcal{G}_{\mu\nu,\rho}(S)\,.
\end{equation}
The equation of motion is now 
\begin{equation}\label{einsteq}
\mathcal{G}_{\mu\nu,\rho}(S)=0\, . 
\end{equation} 
Since the mass of all propagating modes has been sent  to infinity, we should expect this equation  to propagate no modes.  We now confirm this by an explicit canonical analysis. 

To make a time/space split  we set
\begin{equation}
\mu = \left\{0, i\right\}\,, \qquad i=1,2,3 \,, 
\end{equation}
and parametrize the components of the field $S_{\mu\nu,\rho}$ as follows:
\begin{eqnarray} \label{decS}
S_{ij,k} & = & \varepsilon_{ijl} U_{lk} + 2 \delta_{k[i} \zeta_{j]} \nonumber \,, \\
S_{0i,j} & = & W_{ij} + \varepsilon_{ijk} v_k \nonumber \,, \\
S_{ij,0} & = & -2 \varepsilon_{ijk} v_k \nonumber \,, \\
S_{0i,0} & = & w_i \,.
\end{eqnarray}
In the above decomposition, $U_{ij}$ is a symmetric, traceless 3-tensor, $W_{ij}$ is a symmetric 3-tensor and $\zeta_i$, $v_i$ and $w_i$ are 3-vectors. We will fix the gauge invariance \eqref{gaugetran} by imposing a De Donder-type gauge fixing condition
\begin{equation} \label{dedonder}
\partial_i S_{i\mu,\nu} + \frac12 \partial_\nu S_{\mu i,i} = 0 \,.
\end{equation}
Written out in terms of the fields $U$, $W$, $\zeta$, $v$, $w$, this gauge fixing condition is given by
\begin{eqnarray} \label{gaugecond}
& & \varepsilon_{imn} \partial_m U_{nj} = 0 \,, \nonumber \\
& & \partial_i \zeta_i = 0 \,, \nonumber \\
& & \partial_j W_{ji} - \frac12 \partial_i W = \varepsilon_{imn} \partial_m v_n \,, \nonumber \\
& & \varepsilon_{imn} \partial_m v_n = \frac12 \dot{\zeta}_i \,, \nonumber \\
& & \dot{W} = 2 \partial_i w_i \,,
\end{eqnarray}
with $W= W_{ii}$. Note that the above gauge conditions represent 5, 1, 3, 2 and 1 independent conditions respectively. The 16 gauge transformations \eqref{gaugetran} are reducible. Writing \eqref{gaugetran} in terms of
\begin{equation} \label{xipar}
\xi_{\mu \nu} = \frac12 \left( \Xi_{\mu \nu} - \Lambda_{\mu \nu}\right)\,,
\end{equation}
one sees that there is a gauge invariance for the parameters $\xi_{\mu \nu}$
\begin{equation} \label{redxipar}
\delta \xi_{\mu \nu} = \partial_\mu \xi_\nu + \frac12 \partial_\nu \xi_\mu \,.
\end{equation}
In total there are thus only 16 - 4 = 12 independent gauge transformations $\xi_{\mu \nu}$, that are fixed by the 12 gauge fixing conditions \eqref{gaugecond}.

The first condition of \eqref{gaugecond} already implies that $U_{ij}$ does not propagate any physical modes. As
\begin{equation}
2 \varepsilon_{[i|mn} \partial_{m} U_{n|j]} = \varepsilon_{ijm} \partial_n U_{mn} \,,
\end{equation}
the vanishing of the antisymmetric part of this gauge condition, implies that $\partial_j U_{ji} = 0$. Contracting the first equation of \eqref{gaugecond} with $\partial_k \varepsilon_{kli}$ and using $\partial_j U_{ji} = 0$, leads to
\begin{equation}
\nabla^2 U_{ij} = \partial_m \partial_m U_{ij} = 0 \,,
\end{equation}
and hence $U$ does not propagate any physical degrees of freedom. We can thus set it to zero. Similar conclusions for the other fields can be obtained by analyzing the equations of motion \eqref{einsteq}. Upon using the gauge fixing \eqref{gaugecond}, together with $U_{ij}=0$, the components of the Einstein tensor are given by
\begin{eqnarray} \label{einst}
\mathcal{G}_{ij,k} & = & \delta_{k[i} \ddot{\zeta}_{j]} + 2 \partial_{[i} \dot{W}_{j]k} + \varepsilon_{ijk} \partial_m \dot{v}_m - 3 \varepsilon_{ijl} \partial_l \dot{v}_k \nonumber \\ & & + 2 \delta_{k[i} \nabla^2 w_{j]} - 2 \partial_k \partial_{[i} w_{j]} \,, \nonumber \\
\mathcal{G}_{0i,j} & = & - \partial_{[i} \dot{\zeta}_{j]} - \frac12 \delta_{ij} \nabla^2 W + \nabla^2 W_{ij} + \varepsilon_{ijk} \partial_k (\partial_p v_p) \,, \nonumber \\
\mathcal{G}_{ij,0} & = &  2 \partial_{[i} \dot{\zeta}_{j]} - 2 \varepsilon_{ijk} \partial_k (\partial_p v_p) \,, \nonumber \\
\mathcal{G}_{0i,0} & = & \nabla^2 \zeta_i \,.
\end{eqnarray}
{}From $\mathcal{G}_{0i,0} = 0$, one finds that $\zeta_i$ does not propagate any physical modes either and we will thus also set $\zeta_i = 0$. From $\mathcal{G}_{ij,0}=0$, one then infers that $\partial_i v_i = 0$. Together with $\zeta_i = 0$ and the fourth condition of \eqref{gaugecond}, one can then see that $v_i$ also does not propagate any degrees of freedom and can be set to zero as well. The symmetric part of $\mathcal{G}_{0i,j}=0$ then immediately implies that $W_{ij}$ is non-propagating and can be set to zero. Finally, contracting $\mathcal{G}_{ij,k} = 0$ with $\delta_{jk}$ and using the last condition of \eqref{gaugecond} implies that $w_i$ also does not propagate.

To summarise: the equations of motion of NMDG in the infinite mass limit  do not propagate any physical degrees of freedom, as expected. This result can be viewed as a check on our earlier claim that the general solution of the subsidiary condition $\partial^\mu T_{\mu\nu,\rho}=0$ is $T=G(S)$, since we now see that $G(S)=0$ implies that $S=0$ once all gauge invariances have been fixed. 

\section{Massless limit of NMDG} \label{massless}

We have considered the infinite mass limit of NMDG in subsection \ref{infinite}. Now we consider the opposite limit in which $m^2\to 0$. We see from (\ref{NMDGlag}) that the  fourth-order term dominates as $m^2\to 0$, but to actually take the limit we must first multiply by $m^2$. This gives us the ``pure'' fourth-order Lagrangian
\begin{equation}\label{pure4}
{\mathcal{L}}_{m^2=0} = \mathcal{G}^{\mu\nu,\rho}(S) \mathcal S_{\mu\nu,\rho}(S)\, .
\end{equation}
In addition to the gauge invariances (\ref{gaugetran}) this action has the conformal-type gauge invariance
\begin{equation}\label{confgt}
S_{\mu\nu,\rho}\to S_{\mu\nu,\rho} + \eta_{\rho[\mu} \Omega_{\nu]}\,, 
\end{equation}
although the choice $\Omega_\mu= \partial_\mu \phi$ for some scalar  $\phi$ reproduces the $\Xi$-transformation of (\ref{gaugetran}) in the special case that $\Xi_{\mu\nu} =-\eta_{\mu\nu} \phi$,  so two 1-form parameters $\Omega$ and $\Omega'$ correspond to the same conformal-type gauge transformation if they differ by an exact 1-form.  This conformal-type 
gauge invariance is new because it is broken by the ``mass-term'' (\ref{L2}) that we have now dropped from the NMDG Lagrangian.  As we shall see, the new gauge invariance  leads to a  `disappearance' of the  helicity-1  modes from the spectrum, and hence a discontinuity of the $m\to0$ limit\footnote{There is also a discontinuity in the number of propagated modes in the opposite, infinite-mass, limit, but this can be explained as   as an effect of  decoupling at low energy.}. A discontinuity of this sort was to be expected        because a similar  discontinuity in the number of propagated modes occurs in the massless limit of the 3D NMG, which is what one finds by an application to the Lagrangian (\ref{pure4}) of the reduction/truncation procedure of subsection \ref{subsec}; i.e.~the ``pure''  fourth-order 3D Lagrangian analysed by Deser \cite{Deser:2009hb}.  This  3D model was analysed in a different way  in \cite{Andringa:2009yc}, and here we adapt  this analysis  to the 4D case.

First we rewrite the Lagrangian of (\ref{pure4}) as a second-order Lagrangian by introducing an auxiliary field, as explained in subsection \ref{aux}. This gives us the new, but equivalent, Lagrangian
\begin{equation} \label{NMDGeqlagm0}
\mathcal{L}= f^{\mu\nu,\rho}\, \mathcal{G}_{\mu\nu,\rho}(S) -
\frac{1}{2} \left( f^{\mu\nu,\rho}\, f_{\mu\nu,\rho} - 2 f^\mu \, f_\mu \right) \,.
\end{equation}
Now we use the self-adjointness of the generalized Einstein operator to rewrite this Lagrangian as
\begin{equation} \label{NMDGeqlagm1}
\mathcal{L} = S^{\mu\nu,\rho}\, \mathcal{G}_{\mu\nu,\rho}(f) -
\frac{1}{2} \left( f^{\mu\nu,\rho}\, f_{\mu\nu,\rho} - 2 f^\mu \, f_\mu \right) \,.
\end{equation}
In this form, we see that $S$ is a Lagrange multiplier for  the constraint $\mathcal{G}_{\mu\nu,\rho}(f) =0$, which has the solution
\begin{equation}
f_{\mu\nu,\rho} = \partial_{[\mu} H_{\nu]\rho} + \partial_\rho J_{\mu\nu} - \partial_{[\mu} J_{\nu]\rho} \,,
\end{equation}
for some symmetric tensor  $H$ and antisymmetric tensor $J$. Substituting this solution for $f$ (which is a procedure that can be justified by viewing it as integration over $S$ in the path-integral) one finds the new, but equivalent, action
\begin{equation}
\mathcal{L}' = -\frac{1}{2} H^{\mu\nu} \mathcal{G}_{\mu\nu}(H) - \frac{1}{4} F^{\mu\nu\rho}F_{\mu\nu\rho} \,,
\end{equation}
where  $\mathcal{G}_{\mu\nu}(H)$ is the linearized Einstein tensor
\begin{equation}
\mathcal{G}_{\mu\nu}(H) = - \frac{1}{2} \left(\Box H_{\mu\nu} -2\partial_{(\mu} H_{\nu)} + \partial_\mu\partial_\nu H\right)
+ \frac{1}{2}\eta_{\mu\nu}\left( \Box H - \partial^\rho H_\rho\right)\, ,
\end{equation}
with $H_\mu =\partial^\nu H_{\mu\nu}$ and $H = \eta^{\mu\nu} H_{\mu\nu}$, and $F_{\mu\nu\rho}$ is the 3-form field-strength tensor
\begin{equation}
F_{\mu\nu\rho} = 3 \partial_{[\mu} J_{\nu\rho]}\, .
\end{equation}
This result shows that $H$ and $J$ are subject to the gauge-invariances
\begin{equation}
H_{\mu\nu} \to H_{\mu\nu} + \partial_{(\mu} \xi_{\nu)}\, , \qquad
J_{\mu\nu} \to J_{\mu\nu} +  \partial_{[\mu} \zeta_{\nu]} \,,
\end{equation}
for arbitrary vector parameters $(\xi,\zeta)$. Notice that the $f$-tensor is itself invariant under the particular gauge transformation for which $\xi=-3\zeta$.

The essential point  here is that the massless limit of  (linearized) NMDG, defined by the action \eqref{pure4},  is  ghost-free and propagates a massless scalar mode (represented by the 2-form gauge potential $J$) in addition to the two massless graviton modes propagated by the linearized Einstein-Hilbert action for $H$, but there are {\it no massless spin-1 modes}.   We shall now verify this conclusion by a canonical analysis. 

\subsection{Canonical analysis}

We shall use again the parametrization \eqref{decS} of the space/time components of $S_{\mu \nu, \rho}$. Apart from its invariance under  the gauge transformations \eqref{gaugetran}, 
the Lagrangian \eqref{pure4} is also invariant under the conformal transformations \eqref{confgt}. We shall begin by
fixing these conformal transformations. A suitable gauge fixing condition is given by
\begin{eqnarray} \label{confgauge}
S_{\mu i, i} = 0  \qquad \Leftrightarrow \qquad \zeta_i = W = 0  \,.
\end{eqnarray}
As the gauge transformations \eqref{gaugetran} do not preserve this choice of conformal gauge, they must be  augmented by  a compensating conformal transformation. The components of the  parameter $\Omega_\mu$ of this compensating transformation are
\begin{eqnarray}\label{compomega}
\Omega_0 & = &
\frac{2}{3}\left(
    \dot \xi_{ii}
    -2\partial_i\xi_{0i}
    +\partial_i\xi_{i0}
    \right) \,, \nonumber \\
\Omega_i
& = &
\partial_i\xi_{jj}
-2\partial_j\xi_{ij}
+\partial_j\xi_{ji}
\,,
\end{eqnarray}
where $\xi_{\mu\nu}$ is defined in \eqref{xipar}. The compensated $\xi$-transformations are then given explicitly on the fields $U_{ij}$, $W_{ij}$, $\zeta_i$, 
$v_i$, $w_i$ by the following expressions:
\begin{eqnarray} \label{compgauge}
\delta U_{ij}
&=&
\varepsilon_{(i|mn} \partial_m \xi_{n|j)} - \varepsilon_{(i|mn} \partial_{j)} \xi_{mn} \,, \nonumber \\
\delta \zeta_i
&=&
0 \,, \nonumber \\
\delta W_{ij}
&=&
\left(
    \partial_0\xi_{(ij)}
    -2\partial_{(i}\xi_{0j)}
    +\partial_{(i}\xi_{j)0}
    \right)
-\frac{1}{3}\delta_{ij}\left(
    \partial_0\xi_{kk}
    -2\partial_k\xi_{0k}
    +\partial_k\xi_{k0}
    \right) \,, \nonumber \\
    \delta v_k
&=&
-\frac{1}{2}\epsilon_{ijk}\left(
    \partial_i\xi_{j0}
    -\dot\xi_{ij}
    \right) \,, \nonumber \\
\delta w_i
&= &
-\left(
    \partial_i \xi_{00}
    -2\partial_0 \xi_{i0}
    +\partial_0\xi_{0i}
    \right)
-\frac{1}{2}\left(
    \partial_i\xi_{jj}
    -2\partial_j\xi_{ij}
    +\partial_j\xi_{ji}
    \right)\,.
\end{eqnarray}
These gauge transformations are reducible; they still feature the gauge invariance  \eqref{redxipar} of  the parameter $\xi_{\mu \nu}$. Moreover, this parameter is now subject to an extra conformal gauge transformation with  scalar parameter $\Lambda$:
\begin{eqnarray}
\delta \xi_{\mu \nu} = \eta_{\mu \nu} \Lambda \,.
\end{eqnarray}
In total there are now $16-5 = 11$ independent $\xi$-transformations. These can be fixed by the following 11 gauge fixing conditions
\begin{eqnarray} \label{confdedonder}
\epsilon_{imn}\partial_m U_{nj} &=& 0 \,, \nonumber \\
\partial_i W_{ij} & = & 0 \,, \nonumber
\\
\epsilon_{imn}\partial_m v_n & = & 0\,, \nonumber \\
\partial_i w_i  & = & 0 \,.
\end{eqnarray}
As in the infinite mass case, the first of these implies that $U_{ij}$ is non-propagating and can be set to zero. The components of the $\mathcal{C}$-tensor can then be calculated, subject to the gauge conditions \eqref{confgauge} and \eqref{confdedonder} and $U_{ij}=0$. Solving the third condition of \eqref{confdedonder} by
\begin{equation}
v_i = \partial_i \phi \,,
\end{equation}
we find
\begin{eqnarray}
\mathcal{C}_{ij,k} & = & \varepsilon_{ijk} \nabla^2 \Box \dot{\phi} - 3 \varepsilon_{ijl} \partial_k \partial_l \Box \dot{\phi} + 2 \partial_{[i} \Box \dot{W}_{j]k} + 2 \delta_{k[i} \Box \nabla^2 w_{j]} - 2 \partial_k \partial_{[i} \Box w_{j]} \nonumber \\ & & + \partial_k \partial_{[i} \nabla^2 w_{j]} - \delta_{k[i} \Box \nabla^2 w_{j]} \,, \nonumber \\
\mathcal{C}_{0i,j} & = & \Box \nabla^2 W_{ij} + \varepsilon_{ijk} \partial_k \nabla^2 \Box \phi + \frac12 \partial_j \nabla^2 \dot{w}_i \,, \nonumber \\
\mathcal{C}_{ij,0} & = & -2  \varepsilon_{ijk} \partial_k \nabla^2 \Box \phi + \partial_{[i} \nabla^2 \dot{w}_{j]} \,, \nonumber \\
\mathcal{C}_{0i,0} & = & \frac12 (\nabla^2)^2 w_i \,.
\end{eqnarray}
{}From $\mathcal{C}_{0i,0}=0$, it then follows that $w_i$ is non-propagating and can be put to zero. From $\mathcal{C}_{ij,0}=0$, one then gets that $\phi$ obeys the massless Klein-Gordon equation
\begin{equation}
\Box \phi = 0 \,.
\end{equation}
Similarly, from the symmetric part of $\mathcal{C}_{0i,j}=0$, one finds that $W_{ij}$ obeys the massless wave equation
\begin{equation}\label{TT}
\Box W_{ij}^{TT}= 0 \,, 
\end{equation}
where the superscript $TT$ indicates that the symmetric tensor $W_{ij}$ is ``transverse traceless'', i.e. traceless and subject to the gauge condition $\partial_i W_{ij} = 0$, which implies that $W_{ij}^{TT}$ has two independent components,  which are propagated as two helicity-2 modes. Therefore,  there is a total of three propagating modes, two of spin 2 propagated by $W_{ij}^{TT}$ and the other propagated by the scalar $\phi$. 

To show that these modes are propagated unitarily we must return to the Lagrangian  \eqref{pure4}  and express it in terms of the  space/time components of $S$ appearing in 
\eqref{decS}.  Upon imposing the gauge fixing conditions \eqref{confgauge} and \eqref{confdedonder}, we find that
\begin{equation}
{\mathcal{L}}_{m^2=0} = \nabla^2 w_i \nabla^2 w_i
+ 12 (\nabla^2 \phi)\Box (\nabla^2 \phi)
-2W^{TT}_{ij}\nabla^2 \Box W^{TT}_{ij} \,.
\end{equation}
We thus confirm that there are three modes, propagated by $\phi$ and $W^{TT}_{ij}$. Moreover, as $\nabla^2$ is a negative definite operator, we also confirm that 
the kinetic terms for these fields are positive, so that all three modes are propagated unitarily; i.e. they are not ``ghosts''.

\subsection{Another massless limit}

We will conclude this section by showing how the massless limit can be taken in another way that avoids any discontinuity in the 
number of propagating modes. We begin by returning to (\ref{NMDGlag}) and making the field redefinition
\begin{equation}
S_{\mu\nu,\rho} \to \tilde S_{\mu\nu,\rho} = S_{\mu\nu\rho} +  m^{-1} \eta_{\rho[\mu} A_{\nu]}\, . 
\end{equation}
This has no  effect on the quartic term in the action  because of its conformal-type gauge invariance, so the Lagrangian is now 
\begin{equation} \label{NMDGlagA}
\mathcal{L}_{\mathrm{NMDG}} =- \frac12 \tilde S^{\mu\nu,\rho}\, \mathcal{G}_{\mu\nu,\rho}(\tilde S) +
\frac{1}{2m^2} S^{\mu\nu,\rho}\, \mathcal{C}_{\mu\nu,\rho}(S)\,. 
\end{equation}
Although this depends on an additional field $A$, in comparison to the original NMDG Lagrangian, it also has 
an additional gauge invariance: it is invariant under the conformal-type transformation
\begin{equation}
S_{\mu\nu,\rho}\to S_{\mu\nu,\rho} + \eta_{\rho[\mu} \Omega_{\nu]}\,, \qquad A_\mu \to A_\mu -  m \Omega_\mu \, .  
\end{equation}
This additional gauge invariance allows us to set $A=0$, thereby recovering the original NMDG Lagrangian of (\ref{NMDGlag}). 
In other words, $A$ is  a ``St\"uckelberg field''. 

Although the Lagrangian  (\ref{NMDGlagA}) is equivalent to the NMDG Lagrangian (\ref{NMDGlag}) for non-zero mass, it has a different zero-mass limit. To take this limit we first  multiply by $m^2$, as before. Then,  setting $m=0$ and integrating by parts (omitting boundary terms) we arrive at  the new massless Lagrangian
\begin{equation}\label{pure42}
\tilde {\mathcal{L}}_{m^2=0} = - \frac{1}{4} F^{\mu\nu}F_{\mu\nu} +  \mathcal{G}^{\mu\nu,\rho}(S) \mathcal S_{\mu\nu,\rho}(S)\, , 
\end{equation}
where $F_{\mu\nu}= 2\partial_{[\mu}A_{\nu]}$ is the field-strength for $A$. The conformal-type invariance is not lost in this limit because the conformal transformation of $A$ also goes to zero. As expected, however, there is a residual Maxwell-type 
invariance for $A$, ensuring that it propagates only the two massless spin-1 modes that we previously lost in the massless 
limit. What we have done is to take the massless limit in a way that does not change the total number of gauge invariances, and this removes any discontinuity in the number of propagated modes.

\section{Conclusions}

In this paper  we have constructed a 4D analog of the linearized 3D massive gravity theory known as ``new massive gravity'' (NMG) \cite{Bergshoeff:2009hq}. In this construction, the mass scale arises as a consequence of fourth-order terms in the action (curvature-squared terms in the 3D case). A naive extension of the construction from 3D to 4D fails due to the ghosts implicit in higher-derivative theories for $D>3$.  Nevertheless, we have shown that  there {\it is} an extension  of the NMG construction to $D>3$, at least at the linearized level,  if  the graviton field is in an  ``exotic''  Lorentz  representation,  introduced by Curtright \cite{Curtright:1980yk} in the context of  a ``generalized Fierz-Pauli'' model.  In that context the dual graviton field is not a gauge field but we have shown that it may be used to construct an equivalent 4th-order gauge theory for a spin-2 gauge potential in the same ``exotic'' Lorentz representation. 

Although this construction can be carried out for any spacetime dimension $D\ge4$, we have focused here on the 4D case. In that case, we have found an explicit fourth-order action for a dual gauge potential that propagates the five independent modes of a massive graviton. We have called this  (linearized) ``new massive dual gravity'' (NMDG).  The unitarity of this free-field model is guaranteed by the equivalence of the equations of motion to the FP equations and the fact that all five propagated modes are related by rotational invariance.

In the infinite mass limit, the propagating modes decouple. At the Lagrangian level this can be seen from the fact that only a second-order term survives the limit, and this propagates no modes, as confirmed by an explicit  canonical analysis. This is very similar to what happens for 3D NMG; in that case the second-order term  is the Einstein-Hilbert term, which does not propagate any modes in 3D.  

We have also considered the zero-mass limit. This limit can be taken directly in the field equations, and the resulting 
equations can be derived from the ``pure'' 4th-order Lagrangian that results from discarding the 2nd-order term of the NMDG action.  We have shown that the equations of this model propagate the two modes of a massless graviton and an additional massless scalar mode. There are no spin-1 modes however, which means that there is a discontinuity in the number of propagated modes: of  the 5 polarization states at non-zero mass only the 3 even spin states survive at zero mass.  Such a discontinuity was not unexpected because 3D NMG has a similar feature (as do all FP models).  However, we have also shown how it is possible to take the massless limit in a way that does not lead to a discontinuity. In this case the ``missing'' spin-1 modes are propagated by 
an additional Maxwell field that is introduced initially as a St\"uckelberg field for a conformal-type invariance; this can be viewed as an affine version of a spin-2 Higgs mechanism. 

If we suppose that there is some interacting version of the 4D linearized NMDG model constructed here, then our results suggest that it will become equivalent to scalar-tensor GR in the short-distance limit  (possibly with an additional Maxwell field) but will describe massive gravity in the long-distance limit. This could be of relevance to the construction of ``massive gravity'' models that could replace General Relativity on cosmological length scales.  Of course, the introduction of consistent interactions for  ``exotic'' gauge fields  is problematic. In particular, it might seem very unlikely that there exist consistent interactions for dual descriptions of spin-2 (see \cite{Garcia:1998kp} for an analysis of interactions in Curtright's model).  However, one  interacting theory of this general type has been known for a long time, and can be constructed as follows.  Start with the Palatini-form of the Einstein-Hilbert action,  in which both the metric $g$ and the symmetric  affine connection $\Gamma$ are independent fields, 
and then add a cosmological term, with cosmological constant $\Lambda$:
\begin{equation}
\mathcal{L} = \sqrt{-\det g} \left[ g^{\mu\nu}R_{\mu\nu}(\Gamma) - 2\Lambda \right]\, . 
\end{equation}
By partial integration in the action we can arrange for all derivatives to act on the metric, after which  $\Gamma$ becomes an auxiliary field that  may be trivially eliminated. Its field equation determines it  to be the usual Levi-Civita connection and back-substitution yields the Einstein-Hilbert action plus the cosmological term.  On the other hand we could instead eliminate the metric because its  field equation is
\begin{equation}\label{composite}
g_{\mu\nu} =  \frac{(D-2)}{2\Lambda} R_{(\mu\nu)}(\Gamma)\, . 
\end{equation}
Back substitution gives the ($D$-dimensional) Eddington-Schroedinger Lagrangian, which involves only the independent affine connection:
\begin{equation}
{\cal L}_{ES} \propto  \sqrt{\left|\det R_{(\mu\nu)}(\Gamma)\right|}\, . 
\end{equation}
By this construction, the ES theory of gravity  is equivalent to GR with a cosmological constant, as is  well-known\footnote{It is not entirely clear to us how coupling to external sources is achieved in the ES theory. This is presumably done through the composite metric of (\ref{composite}) and something similar might be possible for NMDG.}. What we wish to emphasise here  is  (i) that  the equivalence  to GR is achieved by a version of  the connection-metric  duality that is relevant to the construction of NMDG, and (ii) that the ES theory is an interacting model for an ``exotic'' field, not unlike the field of NMDG.  There are important differences of course, not the least being the fact that the vacuum of the ES model is (anti) de Sitter space rather than Minkowski space, but the example encourages us to imagine that there could be some interacting theory that has NMDG as its linearized limit. It also suggests that it may be necessary to consider an (anti) de Sitter  background. We leave further investigation of this possibility to the future.

\bigskip

\noindent {\bf Acknowledgements} PKT is grateful, for hospitality,  to the University of Groningen, and to the Isaac Newton Institute for Mathematical Sciences, where he was a participant in the program {\it Mathematics and Applications of Branes in String and M-theory} during the completion of this work.  JJFM would like to thank the University of Groningen for hospitality. The work of JR is supported by the Stichting Fundamenteel Onderzoek der Materie (FOM). The work of JJFM has been supported by the Spanish Ministry of Education FPU grant AP2008-00919.
\vskip .1truecm


\providecommand{\href}[2]{#2}\begingroup\raggedright\endgroup

\end{document}